\documentclass[aps,pre,amsmath,amssymb,twocolumn]{revtex4}
\usepackage[utf8]{inputenc}
\usepackage{graphicx}

\usepackage{hyperref}
\newcommand{\subref}[2]{\hyperref[#1]{\ref*{#1}#2}}

\newcommand{\Tr}{\operatorname{Tr}}
\renewcommand{\Im}{\operatorname{Im}}

\newcommand{\bz}{{\textsc{bz}}}
\newcommand{\cor}{{\textrm{cor}}}

\begin{document}


\title{Ioffe-Regel criterion and viscoelastic properties of amorphous solids}
\author{D.\,A. Conyuh}
\author{Y.\,M. Beltukov}
\email{ybeltukov@gmail.com}
\affiliation{Ioffe Institute, 194021 Saint-Petersburg, Russia}
\date{\today}

\begin{abstract}
We show that viscoelastic effects play a crucial role in the damping of vibrational modes in harmonic amorphous solids. The relaxation of a given plane wave is described by a memory function of a semi-infinite one-dimensions mass-spring chain. The initial vibrational energy spreads from the first site of the chain to infinity. In the beginning of the chain, there is a barrier, which significantly reduces the decay of vibrational energy below the Ioffe-Regel frequency. To obtain the parameters of the chain, we present a numerically stable method, based on the Chebyshev expansion of the local vibrational density of states.
\end{abstract}

\maketitle

\section{Introduction}

Damping of vibrational modes plays a crucial role in the thermal conductivity of amorphous dielectrics (glasses). Low-frequency vibrations are well-defined phonons with a long mean free path. However, in a wide range of temperatures, the heat transfer in glasses is determined by another type of delocalized vibrations, which are known as diffusons~\cite{Allen1993, Allen1999}. The crossover between low-frequency phonons and diffusons at higher frequencies is known as the Ioffe-Regel crossover~\cite{Allen1999,Beltukov-2013}. 

In amorphous solids, the attenuation of plane elastic waves (sound) is governed by multiple mechanisms: scattering on two-level systems~\cite{Anderson-1972, Phillips-1972, Jackle-1972} and soft modes~\cite{Buchenau-1992, Ji-2019}, thermally activated relaxation processes~\cite{Jackle-1976, Tielburger-1992}, and scattering induced by structural and elastic disorder~\cite{Ruocco-1999, Dell-1998, Beltukov-2016-PRE, Beltukov-2018, Gelin-2016}. The last contribution is temperature independent and dominates the attenuation in the THz frequency range~\cite{Ruocco-1999, Damart-2017}.

In the low-frequency range, there are phonons with a well-defined dispersion law $\omega(\mathbf{q})$ and a weak damping $\Gamma(\mathbf{q}) \ll \omega(\mathbf{q})$. In this case, the initial plane wave with the wavevector $\mathbf{q}$ oscillates with the frequency $\omega(\mathbf{q})$ with a slow exponential decay. This attenuation can be described using the damped harmonic oscillator (DHO) model.

However, the damping increases rapidly with increasing the wavevector $\mathbf{q}$. For some wavevector $|\mathbf{q}|=q_c$, the damping becomes comparable to the frequency, $\Gamma(\mathbf{q}) \sim \omega(\mathbf{q})$, which corresponds to the Ioffe-Regel criterion. For $q\gtrsim q_c$, the notion of the dispersion law $\omega(\mathbf{q})$ could not be applied. It was shown that the DHO model can not be used for an accurate analysis of vibrational properties for frequencies above the Ioffe-Regel crossover~\cite{Baldi-2016, Buchenau-2014}. Viscoelastic properties are important to study the high-frequency vibrations of amorphous solids~\cite{Luo-2020, Ranganathan-2017, Lemaitre-2006}.

In the theory of viscoelastic relaxation of liquids, it is known that memory effects are important for the relaxation of density fluctuations~\cite{Hansen-1990}. These memory effects can be presented using Mori continued fraction~\cite{Mori-1965}. In this paper we show that the vibrational relaxation in harmonic amorphous solids can also be described using a general viscoelastic model with some memory function $K(t)$. In terms of vibrations, the continued fraction representation corresponds to a semi-infinite mass-spring chain, which reproduces the same memory effects. We present a stable method to find parameters of the arbitrary number of sites in the mass-spring chain.

Another powerful tool to analyze the general properties of disordered systems is the random matrix theory (RMT). Depending on the inherent symmetry properties of different disordered systems, various random matrix ensembles are used~\cite{Evers-2008}. It was shown that the Wishart ensemble naturally arises in the study of vibrational properties due to the requirement of mechanical stability~\cite{Beltukov-2013,Beltukov2015,Conyuh-arxiv}.
In this paper we apply the RMT to find the memory function and the corresponding representation using the mass-spring chain.

This paper is organized as follows. In Section~\ref{sec:damping} we consider a general viscoelastic relaxation of vibrations in harmonic amorphous solids.
Section~\ref{sec:chain} demonstrates that the relaxation is described by a Green function in the form of continued fraction, which corresponds to a one-dimensional mass-spring chain. In Section~\ref{sec:rmt} we obtain the parameters of the chain in the framework of the RMT. Section~\ref{sec:num} demonstrates that the same approach can be used to analyze any given numerical dynamical matrix. In Section~\ref{sec:disc} we discuss the properties of obtained mass-spring chains and compare them to the Ioffe-Regel crossover.

\section{Viscoelastic damping and the memory function}
\label{sec:damping}

The general equation of motion of a solid near equilibrium position can be written as
\begin{equation}
    |\ddot{u}(t)\rangle = - \hat{M}|u(t)\rangle,   \label{eq:harm}
\end{equation}
where $\hat{M}$ is $N\times N$ dynamical matrix with $N$ being the number of degrees of freedom. $N$-dimensional vector $|u(t)\rangle$ describes the deviation of atoms from the equilibrium position at time $t$. 

To study the relaxation of a plane wave with wavevector $\mathbf{q}$, we can solve Eq.~(\ref{eq:harm}) with initial conditions
\begin{equation}
    |u(0)\rangle = 0, \quad |\dot{u}(0)\rangle = |\mathbf{q}\rangle   \label{eq:ic}
\end{equation}
for any given dynamical matrix $\hat{M}$. The relaxation of the initial plane wave is described by the projection
\begin{equation}
    u_\mathbf{q}(t) = \Big\langle \langle u(t) |\mathbf{q}\rangle  \Big\rangle,
\end{equation}
where the big angle brackets denote the averaging over different realizations of the dynamical matrix $\hat{M}$. Assuming the normalization $\langle \mathbf{q}|\mathbf{q}\rangle=1$, we obtain $u_\mathbf{q}(0) = 0$ and $\dot{u}_\mathbf{q}(0) = v_0$ from Eq.~(\ref{eq:ic}).

In a simplified model, the relaxation of $u_\mathbf{q}(t)$ can be described using the DHO model
\begin{equation}
    m_\mathbf{q}\ddot{u}_\mathbf{q}(t) + \eta_\mathbf{q} \dot{u}_\mathbf{q}(t) + k_\mathbf{q}u_\mathbf{q}(t) = 0, \label{eq:damped}
\end{equation}
where the mass $m_\mathbf{q}$, the damping $\eta_\mathbf{q}$, and the stiffness $k_\mathbf{q}$ may depend on the wavevector $\mathbf{q}$. In terms of the DHO model, the Ioffe-Regel crossover separates weakly decaying long-wave vibrational modes and overdamped short-wave vibrational modes. However, it is important to take into account the frequency dependence of the damping, which results in a nonlocal-in-time viscoelastic equation. We will also take into account that the initial equation (\ref{eq:harm}) is time-reversal and do not have the inherent energy dissipation.

\begin{figure}[t]
    \centering
    \includegraphics[scale=0.55]{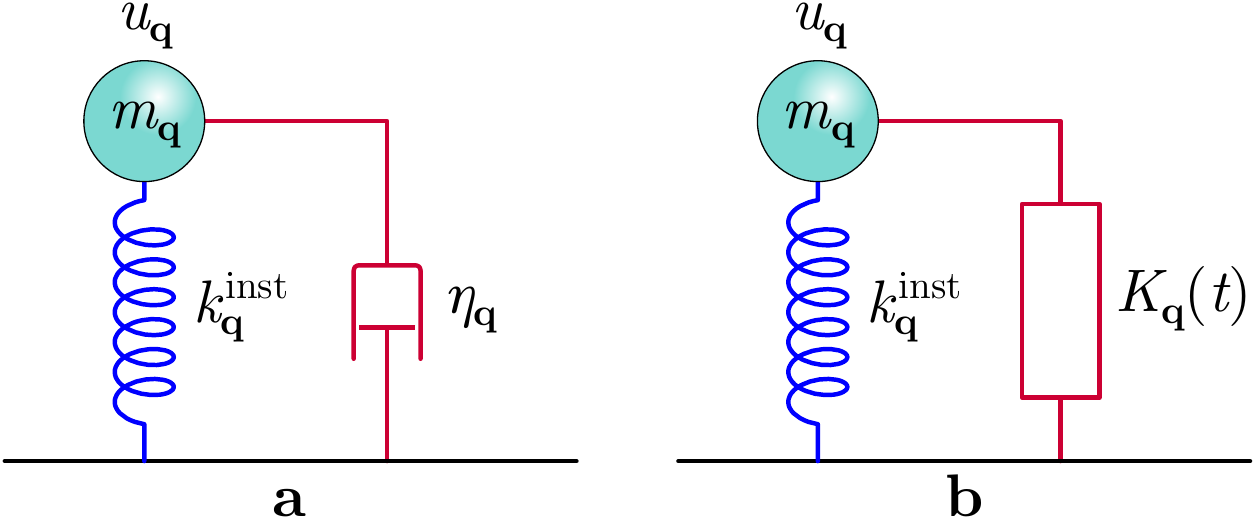}
    \caption{(Color online) a) Damped harmonic oscillator with the mass $m_\mathbf{q}$ (the ball), the stiffness $k^{\rm inst}_\mathbf{q}$ (the spring), and the damping $\eta_\mathbf{q}$ (the dashpot). b) Harmonic oscillator with a general viscoelastic element defined by the memory function $K_\mathbf{q}(t)$ (the rectangle).}
    \label{fig:memory}
\end{figure}

To analyze the relaxation process, we consider the resolvent 
\begin{equation}
    \hat{G}(z) = \left\langle \frac{1}{z-\hat{M}} \right\rangle,
\end{equation}
where $z$ is a complex parameter. The relaxation of a plane wave with initial conditions ${u_\mathbf{q}(0) = 0}$, ${\dot{u}_\mathbf{q}(0) = v_0}$ can be written using the resolvent $\hat{G}(z)$ as (see Appendix~\ref{sec:pw})
\begin{equation}
    u_\mathbf{q}(t) = \frac{1}{2\pi}\int_{-\infty}^\infty\tilde{u}_\mathbf{q}(\omega) e^{i\omega t} d\omega, \label{eq:uq}
\end{equation}
where
\begin{gather}
    \tilde{u}_\mathbf{q}(\omega) = -v_0\mathcal{G}_\mathbf{q}\big((\omega - i0)^2\big), \\
    \mathcal{G}_\mathbf{q}(z) = \langle \mathbf{q}\big|\hat{G}(z)|\mathbf{q}\rangle.
\end{gather}
We can present the Green function  $\mathcal{G}_\mathbf{q}(z)$ as the Stieltjes transform of the spatial Fourier transform of eigenmodes:
\begin{gather}
    \mathcal{G}_\mathbf{q}(z) = \int_0^\infty\frac{{\cal F}_\mathbf{q}(\omega)}{z - \omega^2} d\omega, \label{eq:Green}\\
    {\cal F}_\mathbf{q}(\omega) = \left\langle \sum_n\langle\mathbf{q}|n\rangle\langle n|\mathbf{q}\rangle\delta(\omega - \omega_n) \right\rangle,
\end{gather}
where $|n\rangle$ is $n$-th eigenmode. The Fourier transform of eigenmodes is closely related to the structure factor, which is $S_\mathbf{q}(\omega)=k_BTq^2/(m\omega^2)\mathcal{F}_\mathbf{q}(\omega)$~\cite{Shintani-2008,Beltukov-2016-PRE}. In this paper we use the scalar model to simplify the notations. All qualitative results that we obtain can be applied to the vector model as well. It was shown that vibrations in the scalar and vector models belong to the same class of universality~\cite{Skipetrov-2018}.

In the DHO model with the natural frequency $\omega_\mathbf{q}$ and the frequency-independent damping rate $\Gamma_\mathbf{q}$, we have
\begin{align}
    \mathcal{F}^\textsc{dho}_\mathbf{q}(\omega) &= \frac{2}{\pi}\frac{\omega^2\Gamma_\mathbf{q}}{\big(\omega^2-\omega_\mathbf{q}^2\big)^2+\omega^2\Gamma_\mathbf{q}^2},\\
    \mathcal{G}^\textsc{dho}_\mathbf{q}(z) &= -\frac{1}{\omega_\mathbf{q}^2 - z + \Gamma_\mathbf{q}\sqrt{-z}}.
\end{align}
It corresponds to Eq.~(\ref{eq:damped}) with the stiffness $k_\mathbf{q} = m_\mathbf{q} \omega_\mathbf{q}^2$ and the damping $\eta_\mathbf{q} = m_\mathbf{q} \Gamma_\mathbf{q}$. In the DHO model, the mass $m_\mathbf{q}$ can be chosen arbitrarily.

In the general case, we can present the Green function $\mathcal{G}_\mathbf{q}(z)$ as
\begin{equation}
    \mathcal{G}_\mathbf{q}(z) = -\frac{m_\mathbf{q}}{k^{\rm inst}_\mathbf{q} - m_\mathbf{q}z + \mathcal{G}_{1\mathbf{q}}(z)}.  \label{eq:G_G1}
\end{equation}
It corresponds to the viscoelastic equation of motion
\begin{equation}
    m_\mathbf{q} \ddot{u}_\mathbf{q}(t) + k^{\rm inst}_\mathbf{q}u_\mathbf{q}(t) + \int_{-\infty}^t K_\mathbf{q}(t-t')u_\mathbf{q}(t')dt' = 0 \label{eq:memdamp}
\end{equation}
with the mass $m_\mathbf{q}$, the instantaneous stiffness $k^{\rm inst}_\mathbf{q}$, and the memory function
\begin{equation}
    K_\mathbf{q}(t) = \frac{1}{2\pi}\int_{-\infty}^\infty\mathcal{G}_{1\mathbf{q}}\big((\omega-i0)^2\big) e^{i\omega t} d\omega.   \label{eq:memory}
\end{equation}
In order to find $m_\mathbf{q}$, $k^{\rm inst}_\mathbf{q}$, and $K_\mathbf{q}(t)$, we can present the Green function $\mathcal{G}_{\mathbf{q}}(z)$ as a series
\begin{equation}
    \mathcal{G}_{\mathbf{q}}(z) = \sum_{k=0}^\infty \frac{{\cal F}_{\mathbf{q}}^{(k)}}{z^{k+1}}
\end{equation}
with moments
\begin{equation}
    \mathcal{F}_{\mathbf{q}}^{(k)} = \int_0^\infty \omega^{2k} \mathcal{F}_{\mathbf{q}}(\omega) d\omega.  \label{eq:moments}
\end{equation}
For large $z$, $\mathcal{G}_\mathbf{q}(z)$ is $1/z$ due to the normalization ${\cal F}_{\mathbf{q}}^{\smash{(0)}} = \int_0^\infty{\cal F}_\mathbf{q}(\omega)d\omega=1$. For any harmonic system described by Eq.~(\ref{eq:harm}), all moments $\mathcal{F}_{\mathbf{q}}^{(k)}$ are finite.
In this case, we can assume that $\mathcal{G}_{1\mathbf{q}}(z)$ is also $1/z$ for large $z$. It results in the following values
\begin{align}
    k^{\rm inst}_\mathbf{q} &= m_\mathbf{q} {\cal F}_{\mathbf{q}}^{(1)}, \\ m_\mathbf{q} &= \Big[\mathcal{F}_{\mathbf{q}}^{(2)} - \big(\mathcal{F}_{\mathbf{q}}^{(1)}\big)^2\Big]^{-1}, \\
    \mathcal{G}_{1\mathbf{q}}(z) &= m_{\mathbf{q}}z - k^{\rm inst}_{\mathbf{q}} - \frac{m_{\mathbf{q}}}{\mathcal{G}_{\mathbf{q}}(z)}.    \label{eq:G1}
\end{align}
The decreasing of $\mathcal{G}_{1\mathbf{q}}(z)$ for large $z$ corresponds to the absence of the instantaneous component in the memory function $K_\mathbf{q}(t)$.

\section{Continued fraction and one-dimensional chain}
\label{sec:chain}

We can repeatedly apply the same type of presentation for the Green function:
\begin{equation}
    \mathcal{G}_{n+1,\mathbf{q}}(z) = m_{n\mathbf{q}}z - a_{n\mathbf{q}} - \frac{b_{n\mathbf{q}}^2}{\mathcal{G}_{n\mathbf{q}}(z)}   \label{eq:Gnext}
\end{equation}
with starting Green function $\mathcal{G}_{1\mathbf{q}}(z)$ defined by Eq.~(\ref{eq:G1}). As a result, we obtain the Mori continued fraction~\cite{Mori-1965}:
\begin{align}
    &\mathcal{G}_\mathbf{q}(z) = -\frac{m_{\mathbf{q}}}{k^{\rm inst}_\mathbf{q} - m_{\mathbf{q}}z + \mathcal{G}_{1\mathbf{q}}(z)} \notag \\
    &= -\frac{m_{\mathbf{q}}}{k^{\rm inst}_\mathbf{q} - m_{\mathbf{q}}z - \frac{b_{1\mathbf{q}}^2}{a_{1\mathbf{q}} - m_{1\mathbf{q}}z + \mathcal{G}_{2\mathbf{q}}(z)}} \notag \\
    &= -\frac{m_{\mathbf{q}}}{k^{\rm inst}_\mathbf{q} - m_{\mathbf{q}}z - \frac{b_{1\mathbf{q}^2}}{a_{1\mathbf{q}} - m_{1\mathbf{q}}z - \frac{b_{2\mathbf{q}^2}}{a_{2\mathbf{q}} - m_{2\mathbf{q}}z + \mathcal{G}_{3\mathbf{q}}(z)}}} = \dots   \label{eq:cf}
\end{align}
For each Green function $\mathcal{G}_{n\mathbf{q}}(z)$ we can find the corresponding function $\mathcal{F}_{n\mathbf{q}}(\omega)$, the moments $\mathcal{F}_{n\mathbf{q}}^{(k)}$, and the memory function $K_{n\mathbf{q}}(t)$ using the same relations as in Eqs.~(\ref{eq:Green}), (\ref{eq:memory})--(\ref{eq:moments}). A comprehensive set of relations is given in Appendix~\ref{sec:rel}.

As before, we assume that each Green function $\mathcal{G}_{n\mathbf{q}}(z)$ is $1/z$ for large $z$, which determines the relation between coefficients $a_n$, $b_n$, and $m_n$:
\begin{gather}
    a_{n\mathbf{q}} = m_{n\mathbf{q}} {\cal F}_{n\mathbf{q}}^{(1)}, \quad b_{n\mathbf{q}}^2 = m_{n\mathbf{q}},   \label{eq:an}  \\
    m_{n\mathbf{q}} = \Big[\mathcal{F}_{n\mathbf{q}}^{(2)} - \big(\mathcal{F}_{n\mathbf{q}}^{(1)}\big)^2\Big]^{-1}.  \label{eq:mn}
\end{gather}
The recurrence relation (\ref{eq:Gnext}) can be presented in the form 
\begin{gather}
    \mathcal{F}_{n+1,\mathbf{q}}(\omega) = \frac{m_{n\mathbf{q}} \mathcal{F}_{n\mathbf{q}}(\omega)}{\left|\mathcal{G}_{n\mathbf{q}}\big((\omega-i0)^2\big)\right|^2}.  \label{eq:Fnext}
\end{gather}
The mass $m_{n\mathbf{q}}$ can be defined by the normalization condition ${\cal F}_{n+1,\mathbf{q}}^{(0)} = \int_0^\infty \mathcal{F}_{n+1,\mathbf{q}}(\omega)d\omega=1$, which is equivalent to Eq.~(\ref{eq:mn}). The details of the numerical realization of this recurrence procedure are discussed in Appendix~\ref{sec:Cheb}.

\begin{figure}[t]
    \centering
    \includegraphics[scale=0.55]{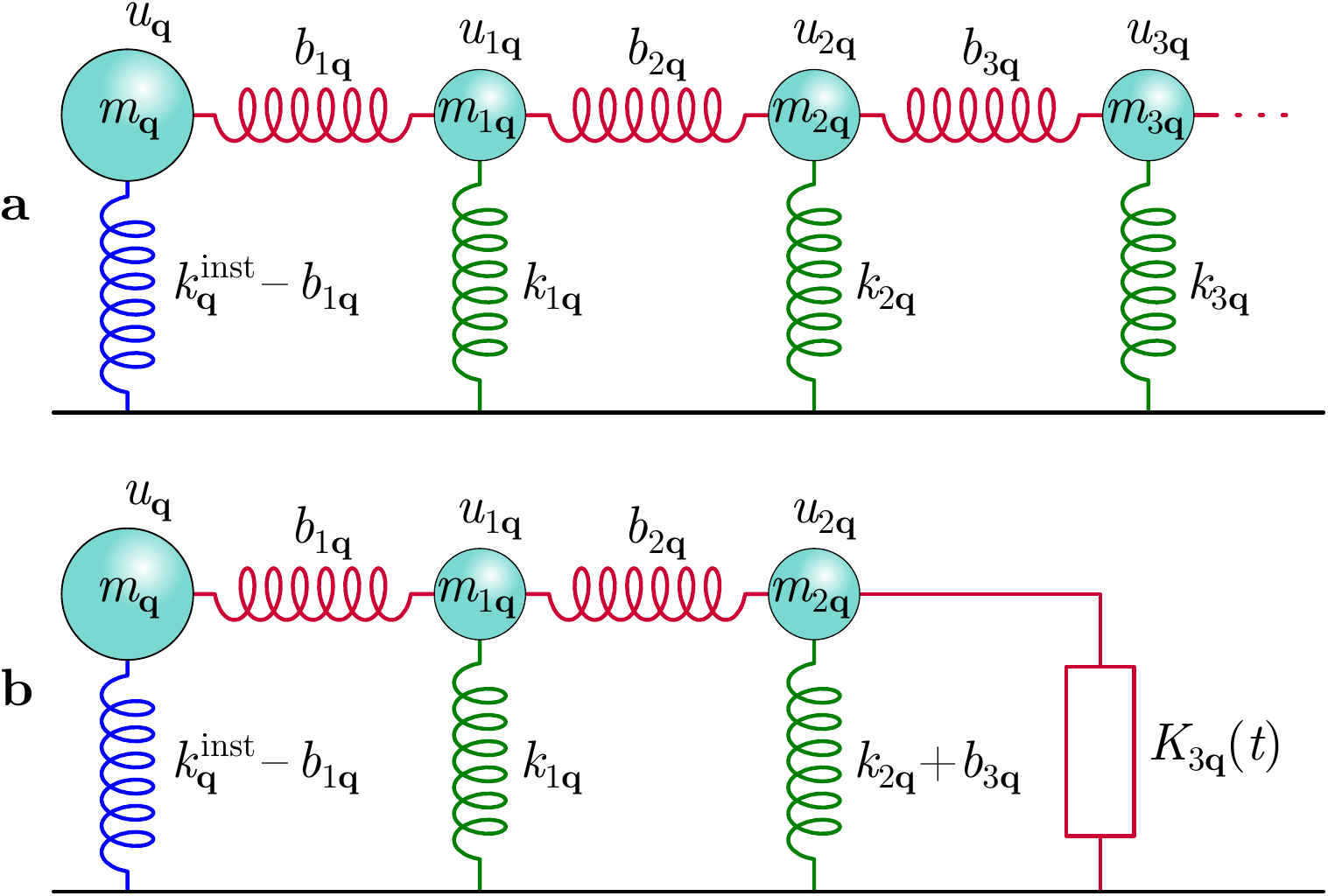}
    \caption{(Color online) a) A general case of the semi-infinite one-dimensional mass-spring model. The displacements of masses $m_\mathbf{q}, m_{1\mathbf{q}}, m_{2\mathbf{q}},\dots$ are denoted by $u_\mathbf{q}, u_{1\mathbf{q}}, u_{2\mathbf{q}},\dots$ respectively. Each spring is denoted by its stiffness. b) Finite representation of the same chain with 3 sites. The tail of the chain is replaced by a general viscoelastic element defined by the memory function $K_{3\mathbf{q}}(t)$ and denoted by the rectangle.}
    \label{fig:chain}
\end{figure}

The continued fraction (\ref{eq:cf}) can be presented using a solution of the following infinite system:
\begin{align}
    m_\mathbf{q}\ddot{u}_\mathbf{q}(t)  &= - k^{\rm inst}_\mathbf{q}u_\mathbf{q}(t) + b_{1\mathbf{q}} u_{1\mathbf{q}}(t), \\
    m_{n\mathbf{q}}\ddot{u}_{n\mathbf{q}}(t) &= -a_{n\mathbf{q}} u_{n\mathbf{q}}(t) + b_{n\mathbf{q}}u_{n-1,\mathbf{q}}(t)  \notag\\
    & \qquad \qquad + b_{n+1,\mathbf{q}} u_{n+1,\mathbf{q}}(t), \quad n \ge 1,  \label{eq:unq_dyn}
\end{align}
where $u_{0\mathbf{q}}(t)\equiv u_\mathbf{q}(t)$ and 
\begin{align}
    u_{n\mathbf{q}}(t) &= \frac{1}{2\pi}\int_{-\infty}^\infty\tilde{u}_{n\mathbf{q}}(\omega) e^{i\omega t} d\omega, \\
    \tilde{u}_{n\mathbf{q}}(\omega) &= -\frac{\tilde{u}_{n-1,\mathbf{q}}(\omega)}{b_{n\mathbf{q}}}\mathcal{G}_{n\mathbf{q}}\big((\omega-i0)^2\big).   \label{eq:unq}
\end{align}
It corresponds to the dynamics of a semi-infinite one-dimensional mass-spring chain (Fig.~\subref{fig:chain}a) with masses $m_\mathbf{q}$ and $m_{n\mathbf{q}}$, horizontal springs with stiffnesses $b_{n\mathbf{q}}$ between consequent masses, and vertical springs with stiffnesses 
\begin{align}
    k_{\mathbf{q}} &= k^{\rm inst}_\mathbf{q} - b_{1\mathbf{q}},\\
    k_{n\mathbf{q}} &= a_{n\mathbf{q}} - b_{n\mathbf{q}} - b_{n+1,\mathbf{q}}
\end{align}
between masses and the common ground. The initial condition of the chain is $u_\mathbf{q}(0) = 0$, $\dot{u}_\mathbf{q}(0) = v_0$, $u_{n\mathbf{q}}(0) = 0$, $\dot{u}_{n\mathbf{q}}(0) = 0$, which defines $u_\mathbf{q}(t)$ for $t>0$.

Using Eq.~(\ref{eq:unq}), the last term in Eq.~(\ref{eq:unq_dyn}) can be replaced by the corresponding memory function, which is known as the Mori-Zwanzig procedure~\cite{Hansen-1990}
\begin{gather}
    b_{n+1,\mathbf{q}} u_{n+1,\mathbf{q}}(t) = -\int_{-\infty}^t K_{n+1,\mathbf{q}}(t-t')u_{n\mathbf{q}}(t')dt'.
\end{gather}
It corresponds to a finite chain shown in Fig.~\subref{fig:chain}b. Therefore, the memory function $K_{n\mathbf{q}}(t)$ describes a response of the tail of the mass-spring chain starting from the site with the number $n$. Using the relations given in Appendix~\ref{sec:rel}, the corresponding function $\mathcal{F}_{n\mathbf{q}}(\omega)$ can be considered as a local vibrational density of states (LVDOS) on site $n$ for the chain under constraint $u_{k\mathbf{q}}(t)=0$ for $k<n$.

\section{A random matrix approach}
\label{sec:rmt}

General vibrational properties could be studied using the random matrix approach~\cite{Conyuh-arxiv}. This approach is based on two main properties of amorphous solids: mechanical stability and the invariance under continuous translation of an amorphous body. In the beginning of this section, we briefly discuss the main points of the random matrix approach.

The mechanical stability of amorphous solids is equivalent to the positive definiteness of the dynamical matrix $\hat{M}$. Any positive definite matrix $\hat{M}$ can be written as $\hat{M} = \hat{A}\hat{A}^T$ and vice versa, $\hat{A}\hat{A}^T$ is positive definite for any (not necessarily square) matrix $\hat{A}$~\cite{Bhatia-2009}. Therefore, we can consider a $N\times K$ random matrix $\hat{A}$ to obtain a mechanically stable system with the dynamical matrix in the form of the Wishart ensemble $\hat{M}=\hat{A}\hat{A}^T$.

Each column of the matrix $\hat{A}$ represents a bond with a positive potential energy $U_k = \frac{1}{2}\big(\sum_{i}A_{ik}u_i\big){}^2$ with $u_i$ being the displacement of $i$-th atom from the equilibrium position~\cite{Beltukov2016,Conyuh-arxiv}. Each row of the matrix $\hat{A}$ corresponds to some degree of freedom. In the random matrix approach, the parameter
\begin{equation}
    \varkappa = \frac{K}{N} - 1
\end{equation}
plays a crucial role. It is a relative difference between the number of bonds $K$ and the number of degrees of freedom $N$. In a stable system with a finite rigidity, the number of bonds should be larger than the number of degrees of freedom, which is known as the Maxwell counting rule. For $0<\varkappa\lesssim 1$, the parameter $\varkappa$ has the same effect as the parameter $z-z_c$ in the jamming transition~\cite{OHern2003}. In real amorphous solids, one can estimate $\varkappa=0.3$ -- 1 depending on the number and the type of covalent bonds~\cite{Conyuh-2020}.

The second important mechanical property of amorphous solid is the invariance under continuous translation. It means that the bond energy $U_k$ should not depend on the shift $u_i \to u_i + const$. Therefore, the matrix $\hat{A}$ obeys the \emph{sum rule} $\sum_i A_{ik} = 0$. It means that the matrix elements $A_{ij}$ are \emph{correlated}. In the minimal model, we can assume that amorphous solid consists of statistically equivalent random bonds. In this case the pairwise correlations between matrix elements $A_{ij}$ can be written as
\begin{equation}
   \langle A_{ik}A_{jl} \rangle = \frac{1}{N}C_{ij}\delta_{kl}, \label{eq:cor}
\end{equation} 
where $\hat{C}$ is some correlation matrix. One can see that the correlation matrix $\hat{C}$ is proportional to the average dynamical matrix: $\hat{C} = \tfrac{N}{K}\big\langle\hat{M}\big\rangle$. For a system with statistically equivalent bonds, the correlation matrix $\smash{\hat{C}}$ is a regular matrix, which describes a lattice with some dispersion law $\omega_\cor(\mathbf{q})$. We assume that there is only one branch of $\omega_\cor(\mathbf{q})$. In the general case, one can apply the summation over different branches below.

Using the random matrix approach, it can be shown that statistical properties of the random matrix $\hat{M}$ are related to the known correlation matrix $\hat{C}$. In the thermodynamic limit $N \to \infty$, there is a fundamental duality relation~\cite{Burda2004}
\begin{equation}
    z\hat{G}(z) = Z\hat{G}_\cor(Z)   \label{eq:duality}
\end{equation}
between resolvents $\hat{G}(z) = \big\langle(z - \hat{M})^{-1}\big\rangle$ and $\hat{G}_\cor(Z) = (Z - \hat{C})^{-1}$ where complex parameters $z$ and $Z$ are related by a conformal map $Z(z)$ defined by the equation
\begin{equation}
    Z\big(\varkappa + 1 + \mathcal{M}_\cor(Z)\big) = z.   \label{eq:map}
\end{equation}
The contour $Z(z)$ in the complex plane for $z=(\omega-i0)^2$ is known as the critical horizon~\cite{Burda-2006}. The moment-generating function $\mathcal{M}_\cor(Z)$ for the correlation matrix $\hat{C}$ is defined as
\begin{equation}
    \mathcal{M}_\cor(Z) =  \frac{Z}{N} \Tr\hat{G}_\cor(Z) - 1 = \sum_{k=1}^{\infty}\frac{\mathcal{M}_{\rm cor}^{(k)}}{Z^k}   \label{eq:Mcor}
\end{equation}
with moments
\begin{equation}
    \mathcal{M}_{\rm cor}^{(k)} = \frac{1}{N} \Tr\hat{C}^k = \frac{1}{V_\bz}\int_\bz \omega_\cor^{2k}(\mathbf{q}) d\mathbf{q}. 
\end{equation}
Here integration is performed over the first Brillouin zone, which has the volume $V_\bz$. The moment-generating function $\mathcal{M}_\cor(Z)$ can also be written as an integral over the first Brillouin zone 
\begin{gather}
    \mathcal{M}_\cor(Z) = \frac{1}{V_\bz}\int_\bz \frac{\omega_\cor^2(\mathbf{q})}{Z - \omega_\cor^2(\mathbf{q})} d\mathbf{q}.   \label{eq:Mcor2}
\end{gather}

\subsection{Relaxation of a plane wave}

Since the correlation matrix $\hat{C}$ is regular, for a given wavevector $\mathbf{q}$ we obtain
\begin{equation}
    \langle\mathbf{q}|\hat{G}_\cor(Z)|\mathbf{q}\rangle = \frac{1}{Z - \omega_\cor^2(\mathbf{q})}.
\end{equation}
Using the duality relation~(\ref{eq:duality}) and the conformal map $Z(z)$ defined by Eq.~(\ref{eq:map}), we can present the Green function $\mathcal{G}_\mathbf{q}(z)$ as
\begin{equation}
    \mathcal{G}_\mathbf{q}(z) = \frac{1}{z - \big(\varkappa + 1 + \mathcal{M}_\cor(Z(z))\big) \omega_\cor^2(\mathbf{q})}.
\end{equation}
For large $z$, we have $Z(z) = (1+\varkappa)z + O(1)$. Therefore, from Eq.~(\ref{eq:Mcor2}), we obtain the asymptotics
\begin{equation}
    \mathcal{M}_\cor(Z(z)) = \frac{(1+\varkappa)\mathcal{M}_{\rm cor}^{(1)}}{z}+O\left(\frac{1}{z^2}\right).
\end{equation}
Therefore, the Green function $\mathcal{G}_\mathbf{q}(z)$ can be presented in the form of Eq.~(\ref{eq:G_G1}) with
\begin{gather}
    k^{\rm inst}_\mathbf{q} = k_{\rm inst} =  1/\mathcal{M}_{\rm cor}^{(1)}, \label{eq:k_inst_rmt}\\
    m_\mathbf{q} = \big[(\varkappa+1)\mathcal{M}_{\rm cor}^{(1)}\omega_{\rm cor}^2(\mathbf{q})\big]^{-1}, \\
    \mathcal{G}_{1\mathbf{q}}(z) = \mathcal{G}_{1}(z) = \frac{\mathcal{M}_\cor(Z(z))}{(1+\varkappa)\mathcal{M}_{\rm cor}^{(1)}}. \label{eq:G1_rmt}
\end{gather}
We omit the subscript $\mathbf{q}$ for the values, which do not depend on the wavevector $\mathbf{q}$. In the framework of the RMT, only the first mass $m_\mathbf{q}$ depends on the wavevector $\mathbf{q}$ in the one-dimensional chain. All other parameters of the one-dimensional chain do not depend on $\mathbf{q}$.

The Green function $\mathcal{G}_1(z)$ corresponds to the LVDOS
\begin{equation}
    \mathcal{F}_1(\omega) = \frac{\omega^2 g(\omega)}{(1+\varkappa)\mathcal{M}_{\rm cor}^{(1)}},  \label{eq:F1}
\end{equation}
where $g(\omega) = \frac{2\omega}{\pi N}\Im\Tr \hat{G}\big((\omega - i0)^2\big)$ is the full vibrational density of states for a given $\varkappa$.
The denominator in Eq.~(\ref{eq:F1}) ensures the normalization of $\mathcal{F}_1(\omega)$.

\begin{figure}[t]
    \centerline{\includegraphics[scale=0.70]{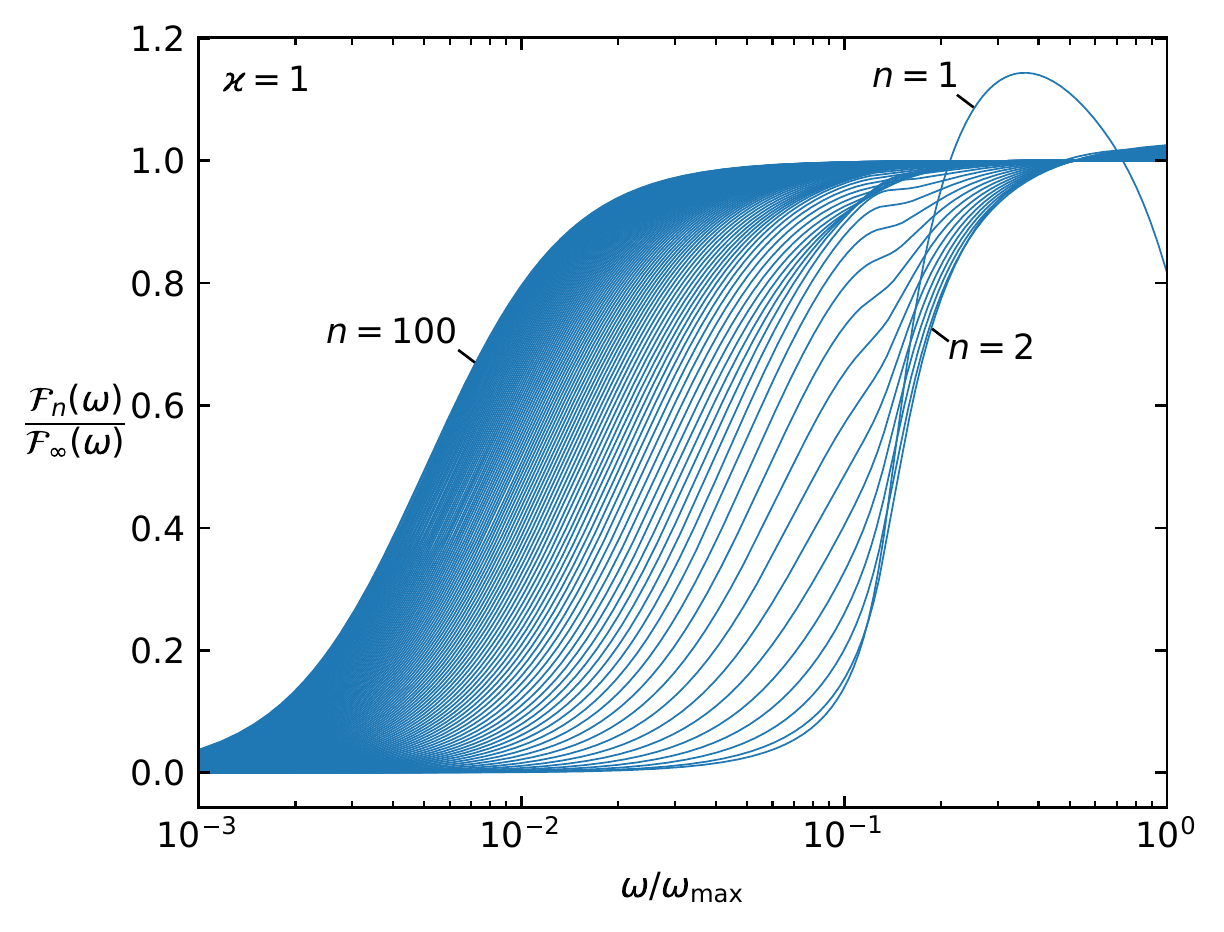}}
    \caption{(Color online) The LVDOS $\mathcal{F}_n(\omega)$ shown as a ratio $\mathcal{F}_n(\omega)/\mathcal{F}_\infty(\omega)$ for $\varkappa=1$ for different values of $n$ (up to $n=100$).}
    \label{fig:Fn}
\end{figure}

\begin{figure}[t]
    \centerline{\includegraphics[scale=0.70]{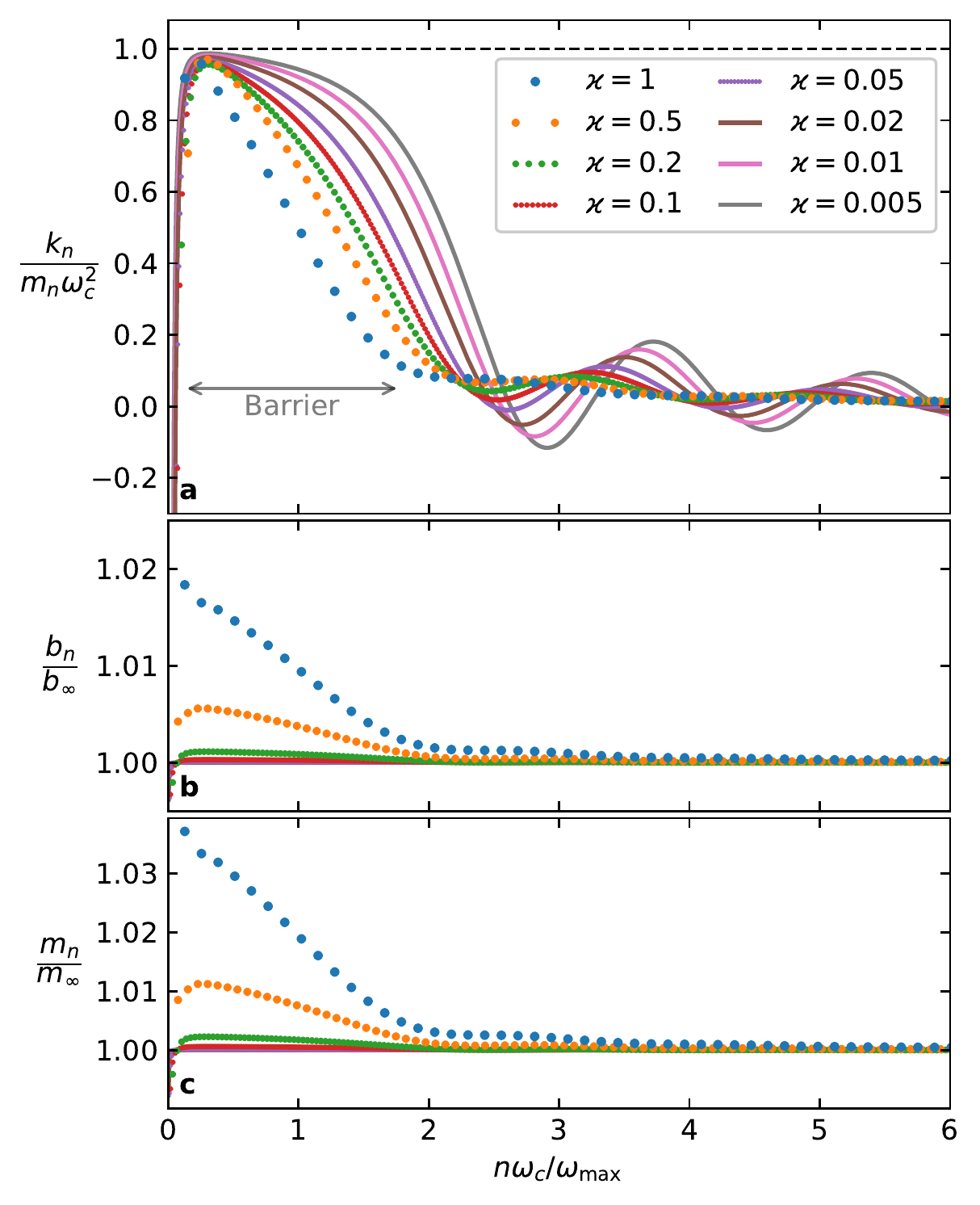}}
    \caption{(Color online) Parameters of the chain as a function of scaled site number $n\omega_c/\omega_{\rm max}$ obtained using the RMT for different values of $\varkappa$. For small values of $\varkappa$, dots merge into lines. In panel (a), the dashed line marks the Ioffe-Regel criterion $k_n/m_n = \omega_c^2$.}
    \label{fig:chain_parameters}
\end{figure}

We can use the recurrence procedure (\ref{eq:an})--(\ref{eq:Fnext}) to obtain the LVDOS $\mathcal{F}_n(\omega)$ and the parameters $a_n$, $b_n$, $m_n$ (all of them do not depend on the wavevector $\mathbf{q}$). For $n\to\infty$, we observe that $\mathcal{F}_n(\omega)$ converges to a stationary solution of Eq.~(\ref{eq:Fnext}):
\begin{equation}
    \mathcal{F}_\infty(\omega) = \frac{16\omega^2}{\pi \omega_{\rm max}^3}\sqrt{1-\frac{\omega^2}{\omega^2_{\rm max}}}.   \label{eq:Finf}
\end{equation}
where $\omega_{\rm max}$ is the maximum frequency in the system. In this case
\begin{gather}
    b_\infty = \frac{a_\infty}{2} = \frac{4}{\omega_{\rm max}^2}, \quad
    m_\infty = \frac{16}{\omega_{\rm max}^4}, \quad
    k_\infty = 0.
\end{gather}

For example, we consider the correlation matrix $\hat{C}$ as a regular matrix on a simple cubic lattice with a unit lattice constant. The non-diagonal elements are $C_{ij}=-1$ if atoms $i$ and $j$ are neighbors and $C_{ij}=0$ otherwise. The diagonal elements are $C_{ii} = 6$. The corresponding dispersion law is
\begin{equation}
    \omega_{\rm cor}^2({\bf q})=4\Big(\sin^2\frac{q_x}{2} + \sin^2\frac{q_y}{2} + \sin^2\frac{q_z}{2}\Big).   \label{eq:cubic_disp}
\end{equation}
In this case, the moment-generating function is $\mathcal{M}_{\rm cor}(Z) = \frac{1}{2}W_s\big(\frac{Z}{2}-3\big)$, where $W_s$ is the third Watson integral~\cite{Zucker-2011}. Using recurrence relation (\ref{eq:Fnext}), we obtain the LVDOS $\mathcal{F}_n(\omega)$ and the values of $a_n$, $b_n$, $m_n$. 

Figure~\ref{fig:Fn} shows $\mathcal{F}_n(\omega)/\mathcal{F}_\infty(\omega)$ for different values of $n$ for $\varkappa=1$. One can observe that $\mathcal{F}_n(\omega)$ is a smooth function, which gradually approaches $\mathcal{F}_\infty(\omega)$ for $n\to\infty$. 

The chain parameters for different values of $\varkappa$ are presented in Fig.~\ref{fig:chain_parameters} as $k_n/(m_n\omega_c^2)$, $m_n/m_\infty$, and $b_n/b_\infty$. The characteristic frequency $\omega_c\sim\varkappa$ is used for the scaling and will be discussed in the next section. One can observe that $k_n/m_n \sim \omega_c^2$ for $n \lesssim \omega_{\rm max}/\omega_c$ and $k_n/m_n \to 0$ for $n \to \infty$. For the better understanding of the behaviour of the obtained mass-spring chain, we consider the low-frequency approximation, which can be analyzed analytically.

\subsection{Low-frequency approximation}

\begin{figure}[t]
    \centering
    \includegraphics[scale=0.55]{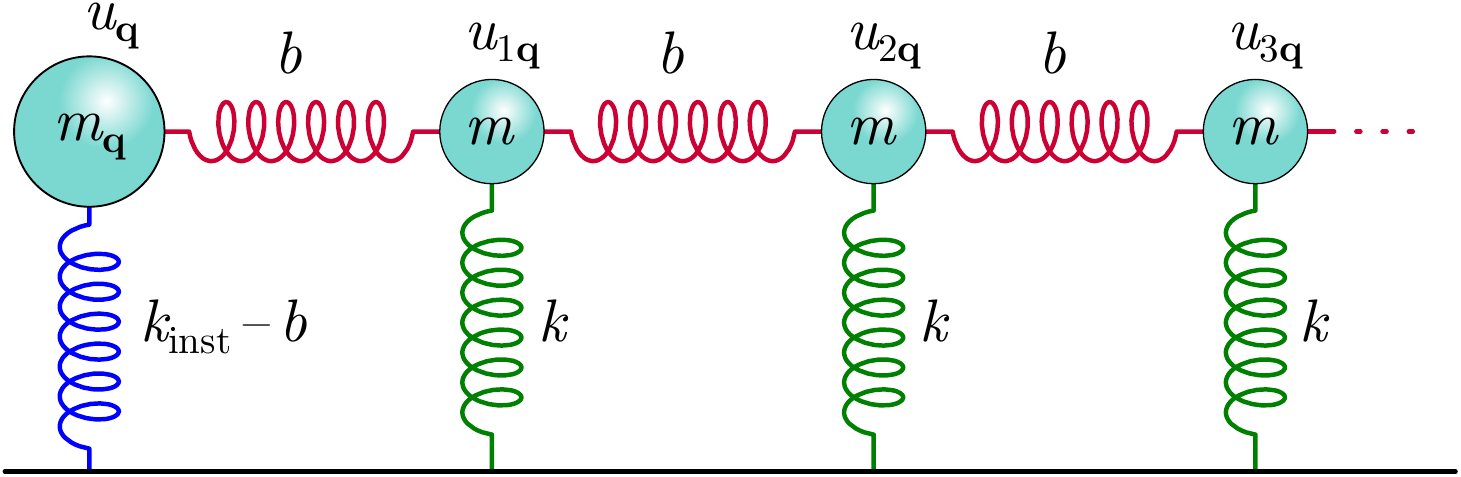}
    \caption{(Color online) A regular one-dimensional mass-spring model, which corresponds to the low-frequency approximation.}
    \label{fig:regular}
\end{figure}

In order to study the low-frequency dynamics, we investigate the behavior of the Green function $\mathcal{G}_\mathbf{q}(z)$ for small values of $z$, which corresponds to small values of $Z$. In this case, the generating function $\mathcal{M}_\cor(Z)$ can be approximated as a linear function of $Z$ for small $Z$:
\begin{equation}
    \mathcal{M}_\cor(Z) = -1 - \beta Z + o(Z),  \label{eq:Mcor_smallZ}
\end{equation}
where
\begin{equation}
    \beta = \frac{1}{V_\bz}\int_\bz \frac{d\mathbf{q}}{\omega_\cor^2(\mathbf{q})}.
\end{equation}
For example, for a simple cubic lattice with the dispersion (\ref{eq:cubic_disp}), the constant is $\beta = w_s/2$ where $w_s = 0.505462$ is the third Watson constant \cite{Zucker-2011}.
For large $Z$, any generating function of the form (\ref{eq:Mcor}) have to decay as $\mathcal{M}_\cor(Z) \sim 1/Z$. Therefore, to study the low-frequency behavior, we can consider the regularized form of Eq.~(\ref{eq:Mcor_smallZ}):
\begin{equation}
    \mathcal{M}_\cor(Z) = \frac{1}{\beta Z - 1},  \label{eq:approx}
\end{equation}
In this case, Eqs.~(\ref{eq:k_inst_rmt})--(\ref{eq:G1_rmt}) becomes
\begin{gather}
    k_{\rm inst} = \beta,\\
    m_\mathbf{q} = \frac{\beta}{(1+\varkappa)\omega_{\rm cor}^{2}(\mathbf{q})}, \\
    \mathcal{G}_{1}(z) = \frac{\beta}{1+\varkappa}\mathcal{M}_\cor(Z(z)).
\end{gather}
Using Eq.~(\ref{eq:map}), we can present $\mathcal{M}_\cor(Z(z))$ in the recurrence form:
\begin{equation}
    \mathcal{M}_\cor(Z(z)) = -\frac{\varkappa+1}{\varkappa + 2 - \beta z + \mathcal{M}_\cor(Z(z))}.   \label{eq:Mcor-rec}
\end{equation}
Thus, we automatically obtain the continued fraction:
\begin{equation}
    \mathcal{G}(z)= -\frac{m_\mathbf{q}}{k_{\rm inst} - m_\mathbf{q} z - \frac{b^2}{a - m z - \frac{b^2}{a - m z - \ldots}}},   \label{eq:uq_cf}
\end{equation}
where  
\begin{gather}
    a = \beta \frac{2 + \varkappa}{1+\varkappa}, \quad b = \frac{\beta}{\sqrt{1 + \varkappa}}, \quad m = \frac{\beta^2}{1+\varkappa}.
\end{gather}
The corresponding mass-spring chain is homogeneous starting from $n=1$, which is shown in Fig.~(\ref{fig:regular}). Each mass in the semi-infinite chain (except the first one) is connected to the ground by a spring with the stiffness $k=a-2b$. It corresponds to the natural frequency 
\begin{equation}
    \omega_c = \sqrt{\frac{k}{m}} = \frac{\sqrt{1+\varkappa} - 1}{\sqrt{\beta}},
\end{equation}
which is the minimal frequency of vibrations, which can propagate along the chain. For small $\varkappa$, we have $\omega_c \approx \varkappa/(2\sqrt{\beta})$. This frequency is used for the scaling in Fig.~\ref{fig:chain_parameters}. The dashed line in Fig.~\subref{fig:chain_parameters}a corresponds to the homogeneous chain obtained in the low-frequency approximation.

Above the frequency $\omega_c$, the chain effectively absorbs the vibrational energy of the first site. Below $\omega_c$, the initial excitation distributes over several sites in the beginning of the chain without any further damping. Indeed, the frequency $\omega_c$ coincides with the Ioffe-Regel frequency in the random matrix approach~\cite{Conyuh-arxiv}. In the low-frequency approximation (\ref{eq:approx}) used in this section, no damping of vibrations below the Ioffe-Regel frequency can be observed. In other words, all frequencies $\omega<\omega_c$ are localized in the beginning of the one-dimensional chain. However, a precise evaluation of the chain parameters shows that $k_n$ gradually decreases to zero for large $n$ (see Fig.~\ref{fig:chain_parameters}). It corresponds to a finite relaxation time for frequencies $\omega<\omega_c$, which will be discussed in Section~\ref{sec:disc}.

From Eq.~(\ref{eq:Mcor-rec}), we can find $\mathcal{M}_\cor(Z(z))$ explicitly:
\begin{multline}
    \mathcal{M}_\cor(Z(z)) = \frac{1}{2}\big(mz - 2b - k \\
    - \sqrt{mz - k}\sqrt{mz - 4b - k}\Big).   \label{eq:explicit}
\end{multline}
Using Eqs.~(\ref{eq:G1_rmt}) and (\ref{eq:memory}), we can find the corresponding memory function $K(t)$ for $k\ll b$ and $\varkappa\ll 1$ (see Appendix~\ref{sec:memory}). In this case the equation of motion (\ref{eq:memdamp}) becomes
\begin{multline}
    m_\mathbf{q} \ddot{u}_\mathbf{q}(t) + \gamma_s \dot{u}_\mathbf{q}(t) + k_s  u_\mathbf{q}(t) \\
    + \int_{-\infty}^t K_l(t-t')u_\mathbf{q}(t')dt' = 0   \label{eq:dyn_low_freq}
\end{multline}
with the short-term damping $\gamma_s=\sqrt{mb}$, the short-term stiffness $k_s = k_{\rm inst} - b - k/2 = \sqrt{kb}$ and the long-term Bessel memory function
\begin{equation}
    K_l(t) = \frac{\sqrt{bk}}{t} J_1\left(t\sqrt{\frac{k}{m}}\right), \quad t>0.   \label{eq:Bessel}
\end{equation}
Equation (\ref{eq:dyn_low_freq}) defines the natural frequency of the first site in the chain
\begin{equation}
    \omega_0(\mathbf{q}) = \sqrt{\frac{k_s}{m_\mathbf{q}}} = \sqrt{\frac{\varkappa}{2}}\omega_{\rm cor}(\mathbf{q}).
\end{equation}
If this frequency is above the Ioffe-Regel frequency $\omega_c$, a strong damping is observed. In terms of the wavevector, it corresponds to $q>q_c$, where the Ioffe-Regel wavenumber is
\begin{equation}
    q_c = \frac{1}{v_{\rm cor}}\sqrt{\frac{\varkappa}{2\beta}}.
\end{equation}
Here we use $\omega_{\rm cor}(\mathbf{q}) = v_{\rm cor}q$ for small $q$. For $q < q_c$, one can find the resonance frequency in Eq.~(\ref{eq:dyn_low_freq})
\begin{equation}
    \omega(\mathbf{q}) = v_{\rm cor}^2 q\sqrt{\beta(2q_c^2-q^2)}.
\end{equation}
In this case, the memory function $K_l(t)$ cancels out the damping term $\gamma_s \dot{u}_\mathbf{q}(t)$ and modifies the resonance frequency: $\omega(\mathbf{q}) > \omega_0(\mathbf{q})$. For $q\ll q_c$, we have $\omega(\mathbf{q}) = \sqrt{2}\omega_0(\mathbf{q})$.

\begin{figure}[t]
    \centerline{\includegraphics[scale=0.70]{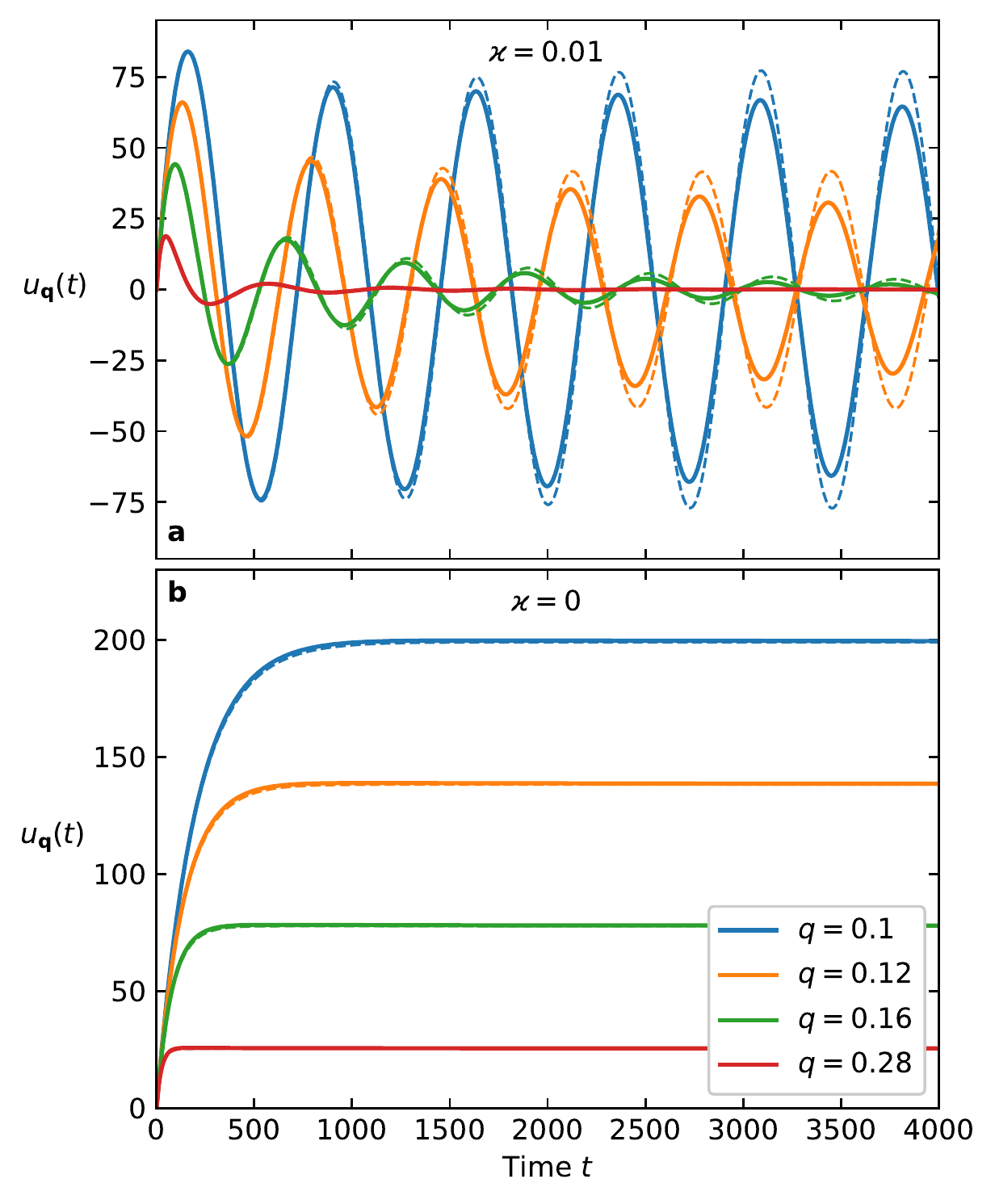}}
    \caption{(Color online) Relaxation of plane waves in the framework of the RMT for $\varkappa=0.01$ (a) and $\varkappa=0$ (b) for the same set of wavenumbers $q$. Solid lines show the precise result, which corresponds to the general chain (Fig.~\ref{fig:chain}). Dashed lines represent a relaxation in the low-frequency approximation, which corresponds to the regular chain (Fig.~\ref{fig:regular}). } 
    \label{fig:relaxation}
\end{figure}

Figure~\ref{fig:relaxation} shows the relaxation of $u_\mathbf{q}(t)$ for a simple cubic lattice with the dispersion $\omega_{\rm cor}(\mathbf{q})$ defined by Eq.~(\ref{eq:cubic_disp}). For $\varkappa=0.01$, the Ioffe-Regel wavenumber is $q_c\approx 0.14$. Figure~\subref{fig:relaxation}a shows the relaxation for two wavenumbers below $q_c$ and two wavenumbers above $q_c$. Solid lines show the exact solution while dashed lines show the low-frequency approximation (\ref{eq:dyn_low_freq}). One can see a good agreement between them. However, for large time $t$ and $q<q_c$, one can observe the slow relaxation of the exact solution while the low-frequency approximation has stationary oscillations.

For $\varkappa=0$, the short-term stiffness $k_s$ and the long-term memory function $K_l(t)$ vanish, and we obtain viscous damping without returning force
\begin{equation}
    m_\mathbf{q} \ddot{u}_\mathbf{q}(t) + \gamma_s \dot{u}_\mathbf{q}(t) = 0.   \label{eq:isostatic}
\end{equation}
Figure~\subref{fig:relaxation}b shows this behavior both for the exact solution and the low-frequency approximation.


\section{A numerical analysis of dynamical matrices}
\label{sec:num}

The same approach can be used to obtain the chain representation for a given dynamical matrix or a given ensemble of dynamical matrices. These dynamical matrices can be obtained by using molecular dynamics simulations or using numerical random matrix models. 

In the previous section, we consider the thermodynamic limit $N\to\infty$ in the framework of the RMT. In this case, there are no fluctuations of vibrational properties. However, in a finite system, fluctuations may be important, especially for the stability of the parameters of the mass-spring chain. In this section, we demonstrate that the proposed recurrence algorithm may be used for a finite numerical system as well.

We use the numerical random matrix model in the form of the correlated Wishart ensemble $\hat{M} = \hat{A}\hat{A}^T$ which can be controlled by the same parameter $\varkappa$. For simplicity, we consider a simple cubic lattice with random bonds and unit lattice constant as in the RMT. However, in contrast to the RMT, the numerical random matrix model has the finite interaction radius. 

For $\varkappa=0$ the matrix $\hat{A}$ is square and the number of bonds $K$ is equal to the number of degrees of freedom $N$. We can consider the following structure of the non-diagonal elements of the matrix $\hat{A}$~\cite{Beltukov-2013,Conyuh-arxiv}:
\begin{equation}
    A_{ij} = \left\{
    \begin{array}{ll}
        \frac{1}{2}\xi_{ij} & \text{if $i$ and $j$ are neighbors}, \\
        0 & \text{otherwise},
    \end{array}
    \right.  \label{eq:A_num}
\end{equation}
where $\xi_{ij}$ are independent Gaussian random numbers with zero mean and unit variance. The diagonal elements are defined using the sum rule $A_{ii} = -\sum_{j\neq i} A_{ji}$. This procedure results in the same correlation matrix $\hat{C}$ that was used in the framework of the RMT in the previous Section.

For $\varkappa>0$ we can use two realizations of square random matrices defined by Eq.~(\ref{eq:A_num}): $\hat{A}^{(0)}$ and $\hat{A}^{(1)}$. The resulting rectangular matrix $\hat{A}$ can be obtained by inserting $\varkappa N$ randomly chosen columns of the matrix $\hat{A}^{(1)}$ into the matrix $\hat{A}^{(0)}$. This random insertion of the new columns corresponds to the random addition of new bonds to the vibrational system. 

\begin{figure}[t]
    \centerline{\includegraphics[scale=0.70]{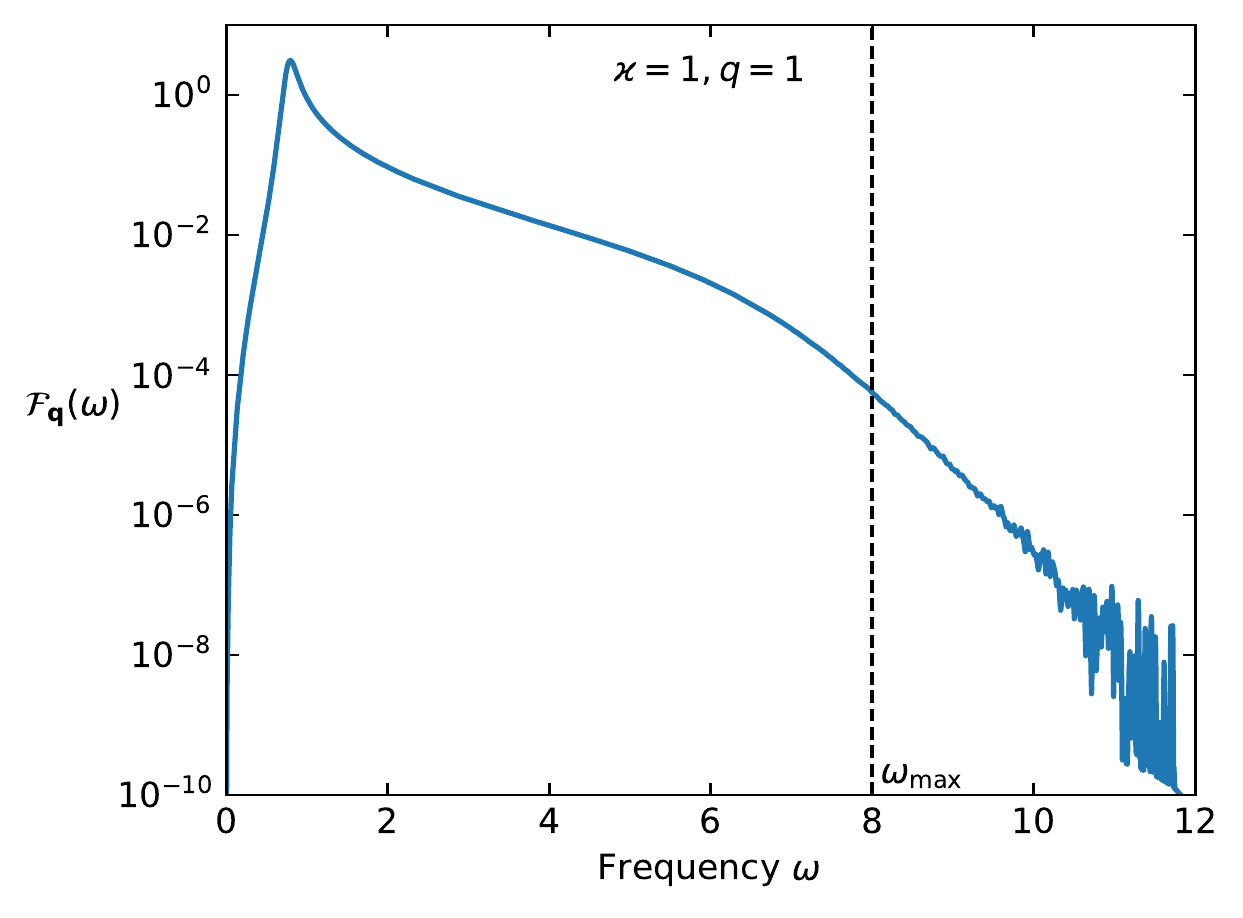}}
    \caption{(Color online) The function $\mathcal{F}_\mathbf{q}(\omega)$ calculated using the KPM for a numerical random matrix with $N=100^3$ atoms, the parameter $\varkappa=1$, and the wave number $q=1$. Averaging over 100 realizations was applied. Vertical dashed line shows the chosen position of $\omega_{\rm max}$.}
    \label{fig:Fnum}
\end{figure}

We use the Kernel Polynomial Method (KPM)~\cite{Beltukov-2016-PRE} to obtain the initial function $\mathcal{F}_{\mathbf{q}}(\omega)$ for the recurrence relation (\ref{eq:Fnext}). The initial function $\mathcal{F}_{\mathbf{q}}(\omega)$ was calculated for a system with $N=100^3$ atoms and different values of the parameter $\varkappa$ and the wavenumber $q$. The averaging over $10^2$ -- $10^3$ realizations was applied. An example for $\varkappa=1$ and $q=1$ is shown in Fig.~\ref{fig:Fnum}.

We use the Chebyshev expansion of $\mathcal{F}_\mathbf{q}(\omega)$ in the frequency range $0\leq \omega \leq \omega_{\rm max}$ to evaluate the recurrence relation (\ref{eq:Fnext}) (see Appendix~\ref{sec:Cheb}).  
The choice of $\omega_{\rm max}$ is important because there is an exponential tail in the high-frequency vibrational density of states without any specific maximum frequency. For any finite system size, there are big relative fluctuations in the high-frequency tail due to a small number of vibrations there (see Fig.~\ref{fig:Fnum}). It may lead to additional fluctuations of the parameters $a_{n\mathbf{q}}$, $b_{n\mathbf{q}}$, $m_{n\mathbf{q}}$ of the obtained chain. Thus, we leave a small number of vibrational modes above $\omega_{\rm max}$ and do not use them in the Chebyshev expansion, which significantly reduces the fluctuations of the obtained parameters.

The KPM is also based on the Chebyshev expansion~\cite{Weisse-2006}. However, the maximum frequency in the KPM, $\omega_{\rm max}^\textsc{kpm}$, should be larger than any frequency in the system for stability purposes. Therefore, we remap the obtained Chebyshev expansion to another one with slightly smaller maximum frequency $\omega_{\rm max}$ to drop a small number of high-frequency modes.

\begin{figure}[t]
    \centerline{\includegraphics[scale=0.70]{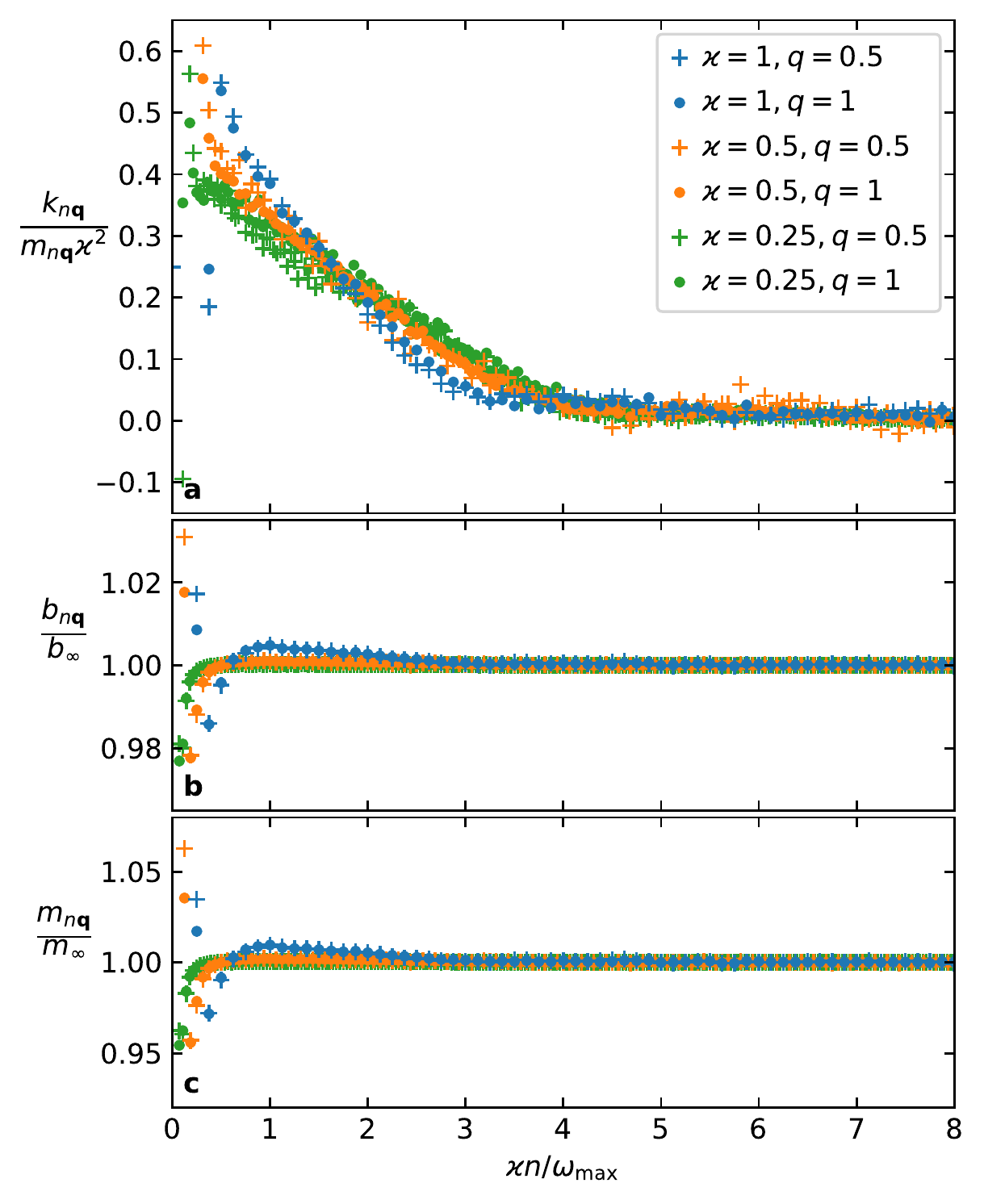}}
    \caption{(Color online) Parameters of the chain as a function of scaled site number $\varkappa n/\omega_{\rm max}$ obtained using the numerical random matrix model (\ref{eq:A_num}) for different values of the parameter $\varkappa$ and the wavenumber $q$.}
    \label{fig:chain_parameters_num}
\end{figure}

The resulting parameters $k_{\mathbf{q}n}$, $b_{\mathbf{q}n}$, $m_{\mathbf{q}n}$ of the mass-spring chain are presented in Fig.~\ref{fig:chain_parameters_num} for different values of the parameter $\varkappa$ and the wavenumber $q$. The results are similar to those obtained in the framework of the RMT (Fig.~\ref{fig:chain_parameters}). In the given scale, all calculated values $k_n/(\varkappa^2m_n)$ almost coincide in Fig.~\subref{fig:chain_parameters_num}a. The masses $m_{\mathbf{q}n}$ and stiffnesses $b_{n\mathbf{q}}$ are close to their stationary values $m_\infty$ and $b_\infty$ respectively. For different wavenumbers $q$, we observe similar parameters of the mass-spring chain.

\section{Discussion}
\label{sec:disc}

We have shown that the relaxation of a plane wave with the wavevector $\mathbf{q}$ coincides with the relaxation of the first site in the semi-infinite one-dimensional mass-spring chain (Fig.~\ref{fig:chain}). The parameters of the chain form the continued fraction representation of the Green function $\mathcal{G}_\mathbf{q}(z)$.

Each site in the chain is connected to the ground by a spring with a stiffness $k_{n\mathbf{q}}$. We observe the same behavior of $k_{n\mathbf{q}}$ in both RMT and numerical analysis of finite random matrices: in the beginning of the chain $k_{n\mathbf{q}}\sim k > 0$ (denoted as a barrier in Fig.~\subref{fig:chain_parameters}a) and then it gradually goes to zero. 

For small values of $\varkappa$, we observe decaying oscillations of $k_{n\mathbf{q}}$. It is known that oscillations of the coefficients in the continued fraction are related to singularities in the density of states~\cite{Gaspard-1973, Hodges-1977}. In our case, the vibrational density of states has a steep behavior near the Ioffe-Regel frequency for small $\varkappa$~\cite{Conyuh-arxiv, Conyuh-2019}. However, these oscillations are not important for qualitative analysis.

For large enough site number $n$, the parameters $a_{n\mathbf{q}}, b_{n\mathbf{q}}, m_{n\mathbf{q}}, k_{n\mathbf{q}}$ converge to their stationary values $a_\infty, b_\infty, m_\infty, k_\infty$, which depend only on the maximum frequency in the system $\omega_{\rm max}$. For electronic systems, it is known, that the coefficients of the continued fraction have stationary values, which depend on the width of the energy band~\cite{Gaspard-1973}. Since $k_\infty=0$, the tail of the mass-spring chain is free and homogeneous. This tail can be considered as a simple thermal bath, which finally absorbs the initial vibrational energy~\cite{Adelman-1976,Li-2007}. At the same time, no explicit damping is introduced in the mass-spring chain, which corresponds to the absence of damping in the initial equation of motion~(\ref{eq:harm}).

In the studied models, all masses $m_{n\mathbf{q}}$ are close to the stationary value $m_\infty$ (except the first one). Thus, the chain can easily absorb the vibrational energy above the Ioffe-Regel frequency $\omega_c=\sqrt{k/m_\infty}$. For frequencies below $\omega_c$, the absorption is much smaller because these frequencies are ``forbidden'' in the beginning of the chain. Depending on the form and the width of the barrier, there is a relatively small absorption below $\omega_c$. The dynamics of the chain can be mapped to a discrete version of a one-dimensional Schr\"odinger equation with the potential energy $k_{n\mathbf{q}}/m_{n\mathbf{q}}$ and the energy $\omega^2$. Therefore, the region with $k_{n\mathbf{q}}/m_{n\mathbf{q}}> \omega^2$ acts as the tunneling barrier (or a high-pass filter).

The first mass in the chain strongly depends on the wavevector $\mathbf{q}$: $m_\mathbf{q} \sim \omega_{\rm cor}^{-2}(\mathbf{q})$. It results in a natural frequency $\omega_0(\mathbf{q}) = \sqrt{\varkappa/2} \omega_{\rm cor}(\mathbf{q})$ of the first site. If this frequency is smaller than the Ioffe-Regel frequency $\omega_c$ for a given wavevector $\mathbf{q}$, then the damping of this vibrational mode is relatively slow. It corresponds to the notion of phonons with well-defined dispersion $\omega(\mathbf{q})$. If $\omega_0(\mathbf{q}) > \omega_c$, a strong damping is observed, which corresponds to the notion of diffusons above the Ioffe-Regel frequency. 

One can note that $\omega_\cor^{2}(\mathbf{q})$ is proportional to the Laplacian on the corresponding lattice. Therefore, in the framework of the RMT, the equation of motion~(\ref{eq:memdamp}) can be rewritten in the real space as
\begin{equation}
    \rho\ddot{u}(\mathbf{r}, t) + k_{\rm inst} \Delta u(\mathbf{r}, t) + \int_{-\infty}^t K(t-t')\Delta u(\mathbf{r}, t')dt' = 0.  \label{eq:spatial}
\end{equation}
This viscoelastic equation is not local in time, but local in space. In a general case, the instantaneous stiffness $k^{\rm inst}_\mathbf{q}$ and the memory function $K_\mathbf{q}(t)$ may depend on the wavevector $\mathbf{q}$ which results in additional spatial convolution in Eq.~(\ref{eq:spatial}). In the low-frequency approximation, from Eq.~(\ref{eq:dyn_low_freq}) we obtain
\begin{multline}
    \rho \ddot{u}(\mathbf{r}, t) + \gamma_s \Delta\dot{u}(\mathbf{r}, t) + k_s  \Delta u(\mathbf{r}, t) \\
    + \int_{-\infty}^t K_l(t-t')\Delta u(\mathbf{r}, t')dt' = 0
\end{multline}
with long-term Bessel memory function $K_l(t)$ defined by Eq.~(\ref{eq:Bessel}). For $\varkappa=0$, from Eq.~(\ref{eq:isostatic}) we obtain a viscous equation without any returning force
\begin{equation}
    \rho \ddot{u}(\mathbf{r}, t) + \gamma_s \Delta\dot{u}(\mathbf{r}, t) = 0.   \label{eq:visc}
\end{equation}
The case $\varkappa=0$ is known as the isostatic state in the jamming transition~\cite{Wyart-2005}. In this case $\omega_c=0$ and the entire low-frequency range is occupied by diffusons. Equation~(\ref{eq:visc}) is consistent with a model of random walks of atomic displacements for isostatic case~\cite{Beltukov-2013}.

The above equation was obtained in the scalar model. In general case, $\mathbf{u}(\mathbf{r}, t)$ is a vector and all constants and memory functions in the above equations become tensors.

The continued fraction presentation (\ref{eq:cf}) is well-known in the classical theory of moments~\cite{Akhiezer-1965}. However, the direct evaluation of moments leads to stability and performance issues for $n \gtrsim 30$. The proposed method is based on the Chebyshev expansion (Appendix~\ref{sec:Cheb}) and shows the numerical stability both for the RMT (Section~\ref{sec:rmt}) and numerical random matrices (Section~\ref{sec:num}). It takes about one hour on a modern computer to find up to $10^5$ coefficients in the continued fraction from the known function $\mathcal{F}_\mathbf{q}(\omega)$. Usually, the most time-consuming part is the calculation of $\mathcal{F}_\mathbf{q}(\omega)$ and its averaging over different realizations of the dynamical matrix $\hat{M}$.

For a given dynamical matrix $\hat{M}$, the continued fraction (\ref{eq:cf}) can be also obtained using the Lanczos method using $|\mathbf{q}\rangle$ as a starting vector~\cite{Dagotto-1994}. However, there are two important issues concerning the Lanczos method. It is known that the Lanczos method is unstable due to the loss of orthogonality. To stabilize the algorithm, an additional reorthogonalization is required, which decreases the performance~\cite{Saad-1992}. The second issue is that the Lanczos method is highly sensitive to a small number of high-frequency eigenmodes. There is no simple way to discard these eigenvalues or average the resulting parameters $a_{n\mathbf{q}}, b_{n\mathbf{q}}, m_{n\mathbf{q}}$ over different realizations of the dynamical matrix $\hat{M}$ in the framework of the Lanczos method. In the proposed algorithm, the averaging is performed directly on $\mathcal{F}_\mathbf{q}(\omega)$ and a small number of high-frequency modes can be easily removed as was discussed in Section~\ref{sec:num}.

\section{Conclusion}

We have shown that viscoelastic relaxation of plane waves in amorphous solids can be considered as dynamics of the one-dimensional semi-infinite chain. The first site in this chain represents the initial plane wave while the rest of the chain represents the memory effects. The initial vibrational energy gradually spreads along the chain, which results in the vibrational relaxation of the initial plane wave.

In the beginning of the chain, there is a natural barrier for frequencies $\omega < \omega_c$, which corresponds to the Ioffe-Regel crossover. In the framework of the RMT, the memory function does not depend on the wavevector $\mathbf{q}$, which results in the viscoelastic equation, which is local in space but not local in time. In the low-frequency approximation, the long-term memory function can be described by the Bessel function.

The proposed method demonstrates numerical stability for studied theoretical and numerical random matrices.

\section{Acknowledgments}
We wish to acknowledge D.\,A.~Parshin and A.\,V.~Shumilin for valuable discussions. The authors thank the Russian Foundation for Basic Research (project no.~19-02-00184) for the financial support.

\appendix

\section{Evolution of plane waves}\label{sec:pw}

For initial conditions $|u(0)\rangle = 0$ and $|\dot{u}(0)\rangle = |\mathbf{q} \rangle$, the solution of (\ref{eq:harm}) is
\begin{equation}
    |u(t)\rangle = v_0 \sum_n |n\rangle \frac{\sin(\omega_n t)}{\omega_n} \langle n | \mathbf{q} \rangle,
\end{equation}
where $|n\rangle$ is $n$-th eigenvector of the matrix $\hat{M}$ and $\omega_n$ is the corresponding eigenfrequency. Therefore, the projection to the plane wave is
\begin{equation}
    u_\mathbf{q}(t) = \left\langle v_0\sum_n \langle \mathbf{q} |n\rangle \frac{\sin(\omega_n t)}{\omega_n} \langle n | \mathbf{q} \rangle \right\rangle.   \label{eq:uq_eval}
\end{equation}
Using the Fourier transform, we can write
\begin{equation}
    u_\mathbf{q}(t) = \frac{1}{2\pi} \int_{-\infty}^\infty\tilde{u}_\mathbf{q}(\omega) e^{i\omega t}d\omega,   \label{eq:uq_t}
\end{equation}
where
\begin{multline}
    \tilde{u}_\mathbf{q}(\omega) = \lim_{\epsilon\to 0^+}\int_0^\infty u_\mathbf{q}(t) e^{-i\omega t - \epsilon t} dt \\
    = \left\langle v_0\lim_{\epsilon\to0^+} \sum_n \langle \mathbf{q} | n \rangle \langle n | \mathbf{q} \rangle \int_0^\infty \frac{\sin(\omega_n t)}{\omega_n} e^{-i\omega t-\epsilon t} dt \right\rangle \\
    = \left\langle v_0\lim_{\epsilon\to0^+} \sum_n \langle \mathbf{q} | n \rangle \frac{1}{\omega_n^2 - (\omega - i\epsilon)^2}\langle n | \mathbf{q} \rangle \right\rangle\\
    = -v_0\big\langle \mathbf{q} \big| \hat{G}\big((\omega - i0)^2\big) \big| \mathbf{q} \big\rangle.
\end{multline}
It is worth to note that $u_\mathbf{q}(t)$ defined by Eq.~(\ref{eq:uq_t}) is zero for $t<0$.

\section{Viscoelastic relations}
\label{sec:rel}
In this Appendix we provide the most important relation between different viscoelastic functions. Relations between the Green function $\mathcal{G}_{n\mathbf{q}}(z)$, the LVDOS $\mathcal{F}_{n\mathbf{q}}(\omega)$, and the moments $\mathcal{F}_{n\mathbf{q}}^{(k)}$ are:
\begin{align}
    \mathcal{F}_{n\mathbf{q}}(\omega) &= \frac{2\omega}{\pi}\Im \mathcal{G}_{n\mathbf{q}}\big((\omega - i0)^2\big), \\
    \mathcal{G}_{n\mathbf{q}}(z) &= \int_0^\infty\frac{{\cal F}_{n\mathbf{q}}(\omega)}{z - \omega^2} d\omega = \sum_{k=0}^\infty \frac{{\cal F}_{n\mathbf{q}}^{(k)}}{z^{k+1}}, \\
    \mathcal{F}_{n\mathbf{q}}^{(k)} &= \int_0^\infty \omega^{2k} \mathcal{F}_{n\mathbf{q}}(\omega) d\omega.\\
\intertext{Relations between the Green function $\mathcal{G}_{n\mathbf{q}}(z)$ and the memory function $K_{n\mathbf{q}}(t)$ are:}
    K_{n\mathbf{q}}(t) &= \frac{1}{2\pi}\int_{-\infty}^\infty\mathcal{G}_{n\mathbf{q}}\big((\omega-i0)^2\big) e^{i\omega t} d\omega,\\
    \mathcal{G}_{n\mathbf{q}}(z) &= \int_{0}^\infty K_{n\mathbf{q}}(t) e^{-t\sqrt{-z}} d\omega.
\intertext{Relations between the LVDOS $\mathcal{F}_{n\mathbf{q}}(\omega)$ and the memory function $K_{n\mathbf{q}}(t)$ are:}
    K_{n\mathbf{q}}(t) &= -\theta(t) \int_0^\infty \frac{\mathcal{F}_{n\mathbf{q}}(\omega)}{\omega}\sin(\omega t) d\omega, \\
    \mathcal{F}_{n\mathbf{q}}(\omega) &= -\frac{2\omega}{\pi} \int_0^\infty K_{n\mathbf{q}}(t)\sin(\omega t) dt,
\end{align}
where $\theta(t)$ is the Heaviside step function. 

\section{Low-frequency memory function}\label{sec:memory}

In the case $k \ll b$, we can write the generating function (\ref{eq:explicit}) in the form
\begin{equation}
    \mathcal{M}_0(Z) = \mathcal{M}_s(Z) + \mathcal{M}_p(Z) + \mathcal{M}_l(Z) + o\left(\frac{k}{b}\right),
\end{equation}
where the main term and two kinds of perturbations have the following form
\begin{align}
    \mathcal{M}_s(Z) &= \frac{1}{2}\left(mz - 2b - \sqrt{mz}\sqrt{mz - 4b}\right), \\
    \mathcal{M}_p(Z) &= \frac{k}{2}\left(\frac{\sqrt{-bmz}}{mz} + \frac{mz-2b}{\sqrt{mz}\sqrt{mz-4b}}\right), \\
    \mathcal{M}_l(Z) &= \sqrt{b(k - m z)}-\sqrt{-bm z}.
\end{align}
Using Eq.~(\ref{eq:memory}) with $z=(\omega - i0)^2$, we obtain the corresponding memory functions
\begin{align}
    K_s(t) &= -\frac{2b}{t} J_2\big(\tilde{t}\big)\theta(t), \\
    K_p(t) &= \frac{k}{2}\sqrt{\frac{b}{m}}\Bigg(J_0\big(\tilde{t}\big)\left[\tilde{t} - \frac{\pi}{2}\tilde{t}H_1\big(\tilde{t}\big)\right]\notag\\
    &\qquad-J_1\big(\tilde{t}\big)\left[2 - \frac{\pi}{2}\tilde{t}H_0\big(\tilde{t}\big)\right]-1\Bigg)\theta(t),\\
    K_l(t) &= \frac{\sqrt{bk}}{t} J_1\left(t\sqrt{\frac{k}{m}}\right)\theta(t),
\end{align}
where $\tilde{t} = 2t\sqrt{b/m}$ is the scaled time, $J_n$ is the Bessel function,  $H_n$ is the Struve function, and $\theta(t)$ is the Heaviside step function. 

The memory function $K_s(t)$ is the main short-term memory function, which is not zero for $k=0$. It coincides with the memory function for the semi-infinite free mass-spring chain~\cite{Li-2007, Kwidzinski-2016}. The memory function $K_p(t)$ is a perturbation of the short-term memory function for nonzero $k$. The memory function $K_l(t)$ is a long-term memory perturbation since it depends on another scaled time $t\sqrt{k/m}$, which scales with $k$.

In the low-frequency approximation, the memory functions $K_s(t)$ and $K_p(t)$ can be considered as an instantaneous response:
\begin{multline}
    \int_{-\infty}^t K_s(t-t')u_\mathbf{q}(t')dt' \\
    \approx \int_{-\infty}^t K_s(t-t')\big[u_\mathbf{q}(t) + (t'-t)\dot{u}_\mathbf{q}(t))\big]dt' \\
    = - b u_\mathbf{q}(t) + \sqrt{mb} \dot{u}_\mathbf{q}(t),
\end{multline}
\vspace{-2\abovedisplayskip}
\begin{multline}
    \int_{-\infty}^t K_p(t-t')u_\mathbf{q}(t')dt' \\
    \approx \int_{-\infty}^t K_p(t-t'))\big[u_\mathbf{q}(t) + (t'-t)\dot{u}_\mathbf{q}(t))\big]dt' = - \frac{k}{2} u_\mathbf{q}(t).
\end{multline}

\section{Chebyshev expansion}
\label{sec:Cheb}

In order to calculate the next LVDOS $\mathcal{F}_{n+1}(\omega)$ using the recurrence relation (\ref{eq:Fnext}), one should calculate ${\mathcal{G}_n\big((\omega - i0)^2\big)}$ from the known LVDOS $\mathcal{F}_n(\omega)$. It can be done using the Chebyshev expansion of the form
\begin{equation}
    \mathcal{F}_n(\omega) = \frac{4\omega^2}{\omega_{\rm max}^3}\sqrt{1 - \frac{\omega^2}{\omega_{\rm max}^2}} \sum_k c_{n,k}U_k\left(\frac{2\omega^2}{\omega_{\rm max}^2} - 1\right),  \label{eq:Cheb}
\end{equation}
where $U_k$ is the Chebyshev polynomial of the second kind. The coefficients $c_{n,k}$ can be obtained using the orthogonal relation:
\begin{equation}
    c_{n,k} = \frac{4}{\pi\omega_{\rm max}^2} \int_0^{\omega_{\rm max}}\!  \mathcal{F}_n(\omega) U_k\left(\frac{2\omega^2}{\omega_{\rm max}^2} - 1\right) d\omega.
\end{equation}
First several moments of $\mathcal{F}_n(\omega)$ are
\begin{align}
    \mathcal{F}_{n}^{(0)} &= \pi c_{n,0} \frac{\omega_{\rm max}^2}{4}, \\
    \mathcal{F}_{n}^{(1)} &= \pi \left(2c_{n,0} + c_{n,1}\right) \frac{\omega_{\rm max}^4}{16}, \\
    \mathcal{F}_{n}^{(2)} &= \pi \left(5c_{n,0} + 4c_{n,1} + c_{n,2}\right) \frac{\omega_{\rm max}^6}{64}.
\end{align}
The normalization condition implies $\mathcal{F}_{n}^{(0)}=1$, which corresponds to $c_{n,0} = 4/(\pi\omega_{\rm max}^2)$. Using Chebyshev expansion (\ref{eq:Cheb}), we can evaluate the Green function:
\begin{multline}
    \mathcal{G}_n\big((\omega - i0)^2\big) = \int_0^\infty\frac{{\cal F}_n(\omega_1)}{(\omega - i0)^2 - \omega_1^2} d\omega_1 \\
    = \pi \sum_k c_{n,k} T_{k + 1}\left(\frac{2\omega^2}{\omega_{\rm max}^2} - 1\right) + \frac{i \pi}{2\omega} \mathcal{F}_n(\omega),   \label{eq:G_Cheb}
\end{multline}
where $T_k(x) = (U_k(x) - U_{k-2}(x))/2$ is the Chebyshev polynomial of the first kind. 

For the stationary LVDOS $\mathcal{F}_\infty(\omega)$ defined in Eq.~(\ref{eq:Finf}), all coefficients $c_{n,k}$ except $c_{n,0}$ are zero. In this case, the Green function has constant absolute value $\big|\mathcal{G}_\infty{\big((\omega-i0)^2\big)}\big| = 4/\omega_{\rm max}^2$ for $0\le\omega\le\omega_{\rm max}$. Thus, for $\mathcal{F}_n(\omega) = \mathcal{F}_\infty(\omega)$ and $m_n=m_\infty$, the recurrence relation (\ref{eq:Fnext}) gives the same LVDOS $\mathcal{F}_{n+1}(\omega) = \mathcal{F}_\infty(\omega)$.

For numerical purposes, we can calculate $\mathcal{F}_n(\omega)$ on a finite number of points, which are known as Chebyshev nodes:
\begin{equation}
    \omega_j = \omega_{\rm max} \sin\left(\pi\frac{2j+1}{4 N_{\rm pts}} \right),
\end{equation}
where $N_{\rm pts}$ is the number of points. In this case, the transformation between $c_{n,k}$ and $\mathcal{F}_n(\omega_j)$ can be performed using the Fast Fourier Transform. The number $N_{\rm pts}$ corresponds to the maximum degree of Chebyshev polynomials that we would take into account.  The number $N_{\rm pts}$ should be several times bigger than the maximum number $n$ used in the recurrence relation.

Using the values on the Chebyshev nodes, the recurrence relation (\ref{eq:Fnext}) can be calculated directly
\begin{align}
    \mathcal{F}'_{n+1}(\omega_j) &= \frac{\mathcal{F}_{n}(\omega_j)}{\left|\mathcal{G}_{n}\big((\omega_j-i0)^2\big)\right|^2},\\
    \mathcal{F}_{n+1}(\omega_j) &= \frac{\mathcal{F}'_{n+1}(\omega_j)}{\mathcal{F}_{n+1}^{\prime(0)}},
\end{align}
producing the following parameters:
\begin{equation}
    m_n = \Big[\mathcal{F}_{n+1}^{\prime(0)}\Big]^{-1}, \quad a_n = m_n \mathcal{F}_{n}^{(1)}, \quad b_n = m_n^2.
\end{equation}
Then we can use again the Chebyshev expansion (\ref{eq:Cheb}) for $\mathcal{F}_{n+1}(\omega)$ and Eq.~(\ref{eq:G_Cheb}) to obtain $\mathcal{G}_{n+1}\big((\omega-i0)^2\big)$.

\bibliography{refs}

\end{document}